# Optimum oxygen content in $La_{0.7}Sr_{0.3}MnO_3$ thin films

Yu M Nikolaenko[1], V N Varyukhin[1], Yu V Medvedev[1], N B Efros[1], I V Zhikharev[1,2], S V Kara-Murza[2] and A A Tikhii[1,2]

[1] Donetsk Institute for Physics & Technology, National Academy of Sciences, Donetsk 83114, Ukraine

[2] Lugansk National University named after T.Shevchenko, Lugansk, Ukraine

E-mail: nik@.fti.dn.ua

**Abstract**

The technological conditions of oxygen content variation in epitaxial $La_{0.7}Sr_{0.3}MnO_{3-\delta}$ films and the peculiarities of change of their transport characteristics were investigated. The films were grown onto single crystal $SrTiO_3$ substrates by an enhanced method of dc magnetron sputtering of a ceramic target. It was revealed, that the oxygen content of the film material with initially strong deviation from stoichiometric composition ($\delta > 0.06$) changes easily in both directions at temperatures $T_p \geq 870$ K, and the result of the oxygen content variation can be well controlled via changes of conductivity minimum temperature ($T_m$). The particularities of indirect accurate control of the oxygen content in the film are described. The existence of oxygen-excessive composition of the film is in principal agreement with the known phase diagrams for a powder samples. It was revealed, the oxygen-excessive content of the film causes the increase of free charge carriers concentration and the decrease of the film conductivity in ferromagnetic state. It is shown, that the oxygen-deficient films can demonstrate the same peculiarity of magnetotransport properties as the films with a deviation from stoichiometric content in cation composition.

PACS: 75.47.Gk, 75.47.Lx, 72.20.-i

## 1. Introduction

The $La_{0.7}Sr_{0.3}MnO_{3-\delta}$ (LSMO) chemical compound is a representative for a solid solution of manganites, which are well known for the effect of colossal magnetoresistance [1,2]. One peculiarity of these materials is a strong correlation between structural, magnetic and electrical properties. As the result, the phase diagrams of these materials are relatively complicated, and interpretations of experimental dependencies are ambiguous. Because LSMO is the most investigated solid solution of manganite with a relatively simple phase diagram, it is an interesting object for the investigation of the physics of these materials and for a quantitative verification of existing theories.

The next peculiarity of manganites is their multicomponent oxide structure. The production of samples with high quality crystalline structure is a separate problem. The preparation technology must provide additionally the stoichiometry with respect to an anion composition and minimum quantity of defect crystalline cells.

Recently we have described the preparation technology of epitaxial films of manganite solid solution with high quality structural and excellent electrical characteristics [3,4]. We have used the method of magnetron sputtering of ceramic targets and provided the conditions for epitaxial growth of atomic layers on the surface of single crystal substrates heated up to 900-1000 K. In order to achieve the stoichiometric cation content, the material from the target is transferred to the substrate by nanosize particles. At the hot substrate the particles are rebuilt into the film atomic layers thank to high surface diffusion. Second very important procedure to provide excellent film characteristics is the thermal treatment of the film structure under oxygen contenting atmosphere. For the real tuning of the oxidation procedure the control of the film oxygen content must be accurate enough. Because such control is not accessible by direct measurement methods, the optimum film oxidation was determined indirectly in this paper.

Earlier [4], a hierarchic scheme for the general estimation of a LSMO film quality was proposed by us. For such estimation we use both the value of temperature minimum of film conductivity ($\sigma_{min}$), and the relation between maximum and minimum conductivity values in the temperature range 100-400K. The parameter $\sigma_{min}$ is very sensitive to a variation of the oxygen index $\delta$, as well as to a deviation from cation stoichiometric composition. For the best LSMO film $\sigma_{min}$ exceeds the threshold value of metal conductivity according to the Mott criterion. The films with excessive contamination of Mn [5] are characterized by approximately one order of magnitude lower values of $\sigma_{min}$.

The understanding of the particularities of the structural self-organization of multicomponent crystals during thermal treatment is of interest for practical applications, and the description of defect formations, when a correlation between the content of different types of point defect is supposed, is of scientific interest.

In this work we described the evolution of transport characteristics of LSMO films in dependence on oxygen content including the vicinity of stoichiometric composition, and peculiarities of film phase diagram in contrast with powder sample's phase diagram [6]. For the control of optimum oxidation of LSMO films with relatively high accuracy the indirect identifiers were revealed.

## 2. Experiment

Our technology of films growth was described in [3]. Figure 1a shows a set of temperature dependencies of LSMO (F641N) films with the thickness 100 nm. The first curve was measured immediately after the film growth without annealing. The numbers of other curves correspond to each measurement after three procedures of the film annealing in air at 870 K and normal pressure, except the second procedure when the pressure was reduced up to $10^{-2}$ atm. The changes in the R(T) curves demonstrate the possibility of a reversible variation of oxygen content in the film. The values of the oxygen index ($\delta$) and the temperatures of the film resistance maximums ($T_m$) are presented in the table 1 together with technological parameters of the thermal treatment procedures. In table 1 $T_t$ is the annealing temperature, $P_t$ – pressure of air, and $\Delta t$ - duration of annealing. The values of $\delta$ were determined with the help of diagram of [7], see also [4].

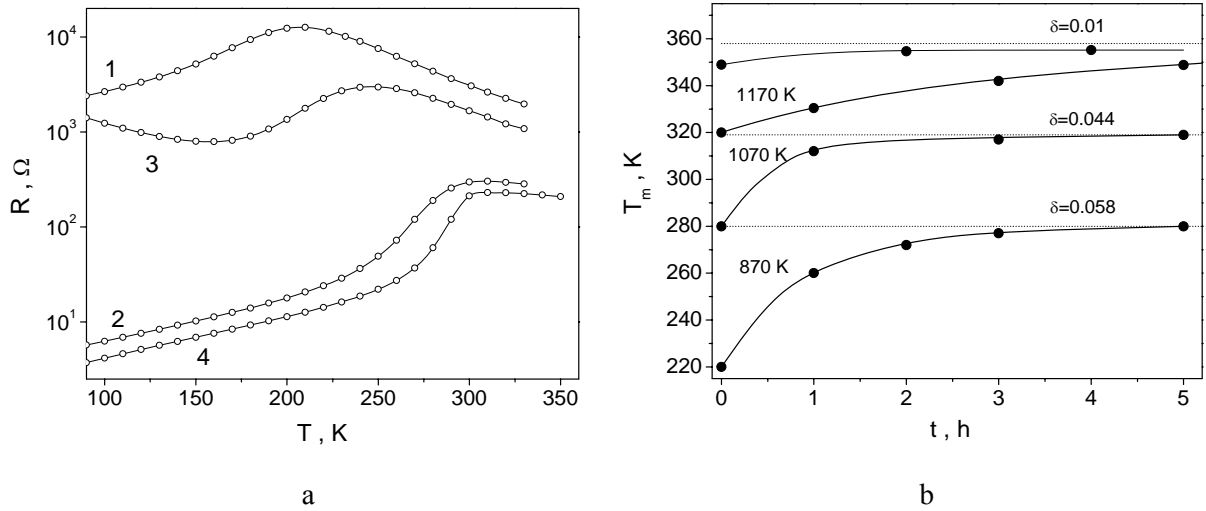

**Figure 1.** (a) - Temperature dependencies of LSMO film resistance before and after 3 procedures of thermal treatment and (b) - Values of maximum resistance temperature in the dependence on duration of thermal treatment procedure at $T_t$=870, 1070 and 1170 K.

**Table 1.** The characteristics of the thermal treatment procedure.

| Number of R(T) dependence | $T_t$ (K) | $P_t$ (atm) | $\Delta t$ (min) | $T_m$ (K) | $\delta$ (r.u.) |
|---|---|---|---|---|---|
| 1 | - | - | - | 210 | 0.08 |
| 2 | 1070 | 1 | 35 | 308 | 0.051 |
| 3 | 1170 | 0.01 | 60 | 246 | 0.071 |
| 4 | 1170 | 1 | 30 | 310 | 0.048 |

As seen from the table 1 data and figure 1a, at a relatively large $\delta$ the film oxygen content is easily varied in both directions. Figure 1b shows the peculiarity of oxygen content variation at several temperatures. The significant fact consists in a saturation of temporal dependence of $\delta(t)$ at fixed temperature. Under the condition of appropriate duration of the thermal treatment procedure (of the order 10 h), the limiting value of oxygen index decreases consecutively from $\delta$=0.058 at $T_t$=870 K to $\delta$=0.01 at annealing temperatures equal to 1170 K. Significant increase of annealing temperature is dangerous, because at $T_t$>1400 K an irreversible decomposition of film material occurs. Figure 2 shows SEM images of fragments of (F641N) film surface. Figure 2a demonstrates the phase homogeneity of the film after annealing at 1170K, and (figure 2b) a destruction of the film after annealing at $T_t$=1470 K. As a result of the film decomposition, separate oxide grains with significant excessive composition of one type of cation (Mn, Sr or La) were formed, see figure 2c.

Using oxygen atmosphere at normal pressure and annealing temperature 1170 K we can obtain oxygen-excessive film composition.

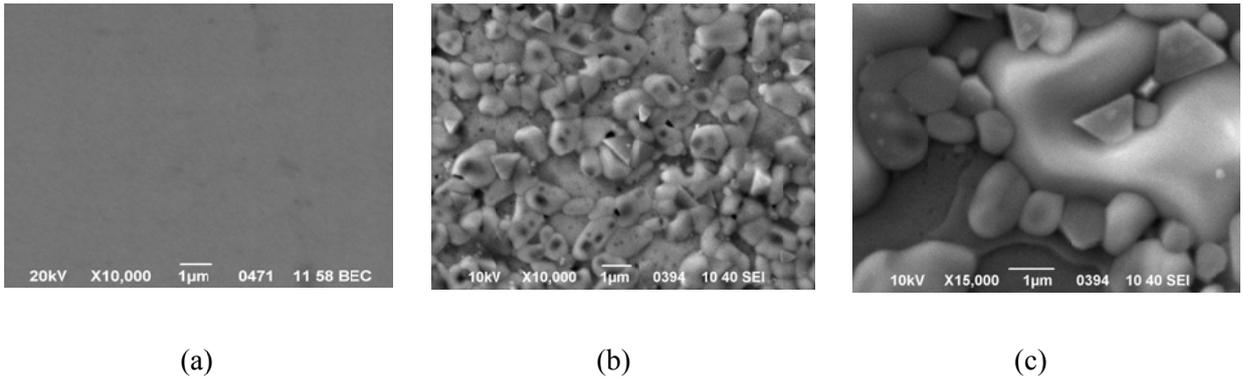

(a) (b) (c)

**Figure 2.** SEM images of fragment of (F641N) film after annealing at T=1170 K (a) and 1470 K (b,c).

Figure 3 shows the temporal variation of the temperature value $T_m$ during multistage thermal treatment procedure. The value of $T_m$ corresponds to the temperature at the minimum of the F64004 film conductivity. The conditions of each step of the annealing procedure are shown in the lower part of figure 3. The main peculiarity of this dependence is the maximum of $T_m(\delta)$ at point t=13h. We attribute this fact with minimal quantity of defect crystalline cells in the film structure, and with maximal nearness to the oxygen stoichiometric composition.

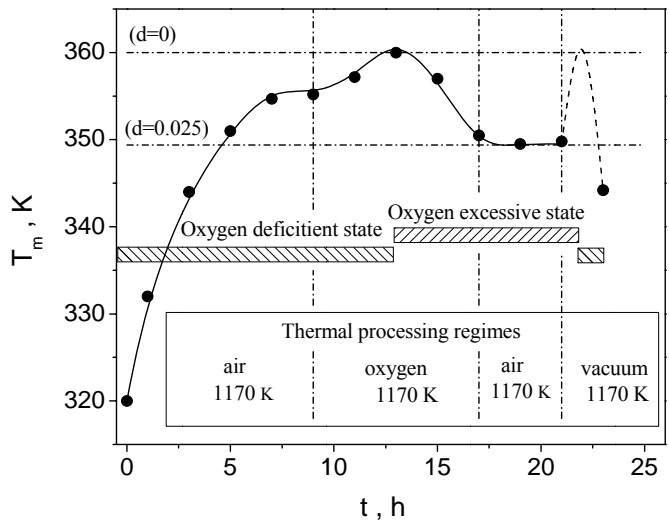

**Figure 3.** The temperature corresponding to the film conductivity minimum as function of the duration of the annealing procedure. The annealing condition is presented in the lower part of the figure.

It is clear, that the presence of defective crystal cells in the film structure impair the formation of a coherent ferromagnetic (FM) state, because the magnetic exchange interaction is very sensitive to the length and angle of Mn-O-Mn bonds [1,2]. The mechanism of atom cell distortion connected with oxygen vacancies at oxygen-deficient composition and with cation vacancies at oxygen-excessive one (this question is discussed later). A decrease of the $T_c$ value for oxygen-excessive compositions of ceramics samples is mentioned in [8].

More complicated is the question of film conductivity for oxygen excessive composition. As a rule [4,7], an oxygen-deficit of LSMO film composition results in decrease of conductivity in both FM and paramagnetic (PM) states (one exception is discussed later). Such behavior can be well explained by the concentration of $Mn^{4+}$ ions and, consequently, a concentration of free charge carriers (holes in $e_{2g}$ band of Mn). The concentration is determined by the parameter $z=x-2\delta$ [7], where x is the composition part of Sr. Oxygen vacancies decrease the concentration of free carriers and their kinetic energy, which in the model of double exchange interaction is the main factor for the existence of the FM state [9].

In the figure 4 two temperature dependencies of the F64004 film are presented. Curve 1 corresponds to the optimum oxygen composition and curve 2 to the oxygen-excessive one. As is seen, the curve 1 has smaller values than curve 2 at low temperature in the FM region and larger ones in the PM region, including some vicinity of $T_m$. This fact indicates that formally the relation for z is valid for oxygen-excessive compositions. I.e. an excessive oxygen atom creates acceptor states and increases the hole concentration. In the PM region this effect increases the film conductivity. In the FM state the conductivity is strongly dependent on the spin disorder scattering. The decrease of the charge carrier mobility is a result of the energy loss to spin direction changing [9,10]. A minimum of mobility appears in the vicinity and above $T_c$. Thereby, we must conclude, that the conductivity decrease of the film with oxygen-excessive compositions in the FM region is caused by a spin order worsening, and this effect completely compensates the increase of the conductivity due to the increase of free charge carriers.

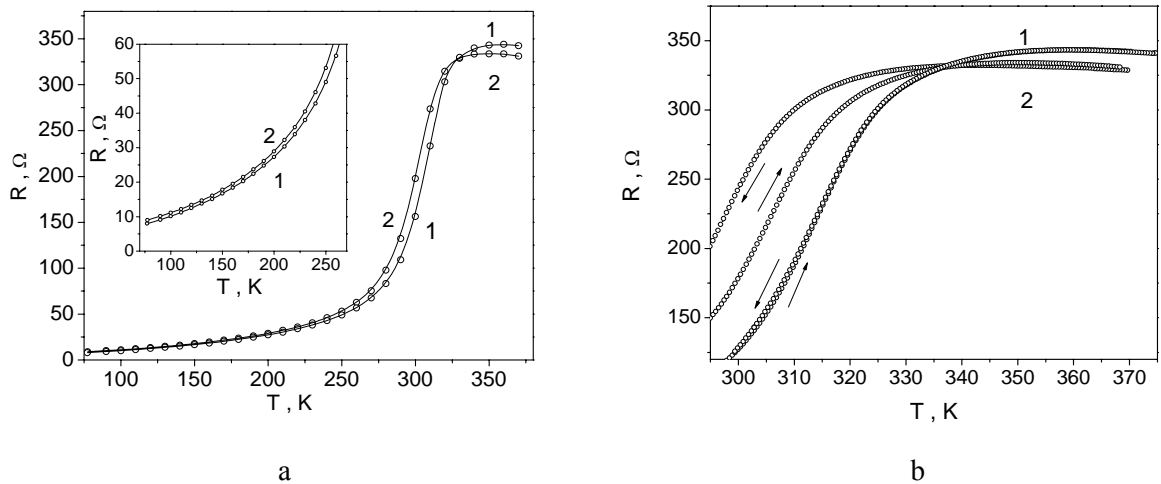

**Figure 4.** (a) Temperature dependencies R(T) of the F64004 film with optimum oxidation (curve 1) and oxygen-excessive composition (curve 2) and (b) Hysteresis of R(T) dependence of film with oxygen-excessive composition and absence of hysteresis at optimal oxygen content. The data was measured in the regime of temperature variation with constant velocity about 0.5K/s.

The curves of figure 4a were measured in the regime of stable temperature with duration of not less than 5 min for each point. In the regime of temperature variation with constant velocity, the R(T) dependence of films with oxygen-excessive composition exhibit the pronounced retarded hysteresis, see figure 4b. We attribute the nature of the hysteresis with a temporary structural relaxation of the film at non-equilibrium conditions in the presence of a film structure defect crystal cells. At optimum oxygen content in the film the hysteresis is absent, see figure 4b. Because an oxygen deficit as well as an oxygen excess causes both a hysteresis of R(T) and a decrease of $T_m$, it is interesting to compare the hysteresis widths of two different states of the film. The experiment showed that at nearly equal $T_m$ in the case of oxygen excessive composition the width of the hysteresis is large significantly. This fact indicates that a difference in the defect structures of the films exists.

Now we focused our attention on a very week variation of the R(T) dependence in the PM region. As can be seen, the maximum of the R(T) curves in figure 4 is very weak. This fact must be traditionally interpreted by the absence of shallow localized states [2,9]. After an increase of the oxygen deficit, a maximum in the temperature dependence of the film resistance becomes pronounced and a dependence R(T) at $T>T_m$ can be well described by an Arrhenius function $R(T) \sim \exp(\Delta E/kT)$. At significant deviation from oxygen stoichiometric composition ($\delta>0.06$) this simple description becomes invalid. The condition $R(T>T_m)$ is better described by $\sigma(T) = \sigma_v(\delta) + \sigma_0 \cdot \exp(-\Delta E/kT)$, see insert of figure 5a. Here $\sigma_v$ is independent from temperature. The value of $\sigma_v$ at $\delta$=const is close to the conductivity minimum.

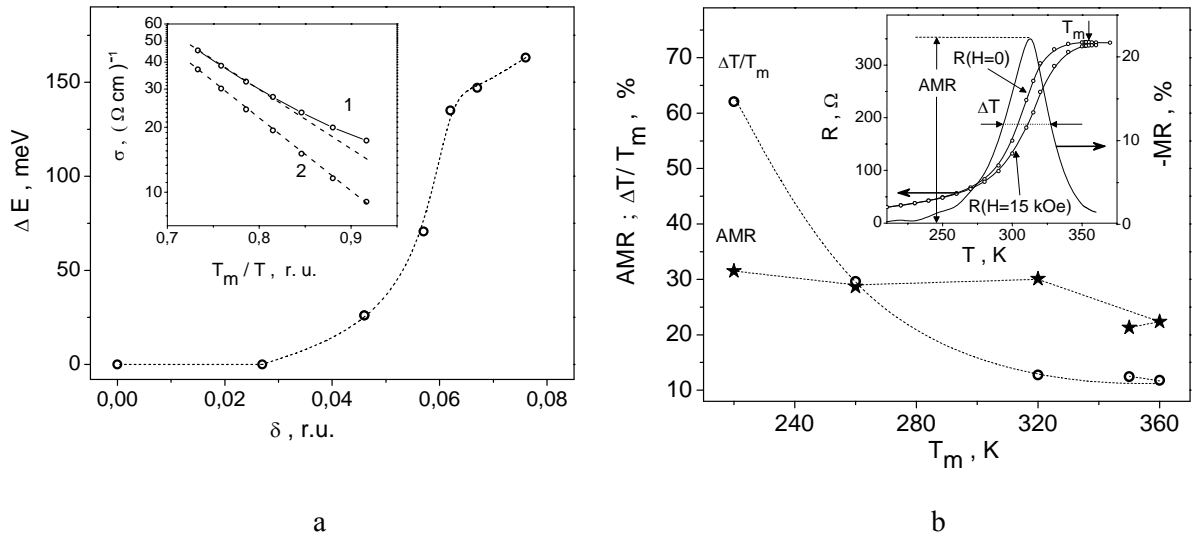

a            b

**Figure 5.** (a) The variation of the energy parameter ΔE as function of oxygen index. Insert: Temperature dependence of F64004 film conductivity σ(T) (curve 1) and σ(T)- $\sigma_v$ (curve 2) in the scale $T_m/T$. (b) Variation of AMR and relative width ($\Delta T/T_m$) of MR(T) curve (see insert of figure 5b) in the dependence on $T_m$ (δ).

The figure 5a shows the variation of the energy parameter ΔE as function of oxygen index for the F64004 film. The zero value of ΔE at δ<0.02 proves the absence of thermal activation of free carriers from localized states. Such a behavior corresponds to heavily doped semiconductors. With the increase of the oxygen index up to δ=0,08, the parameter ΔE reaches 0,16 eV. This confirms, that films with oxygen-deficient compositions correspond to compensated semiconductors. The parameter $\sigma_v$ introduces into account conductance of impurity band, which is accompanied by the increase of oxygen vacancy concentration.

In the model of double exchange interaction, a magnetic field (MF) is considered as a factor for the decrease of spin disorder and spin temperature fluctuations [9,10]. The resistance is a quadratic function of magnetization, which determines the dependence R from temperature and MF. Thus, the magnetoresistance (MR=100%·(R(H,T)-R(0,T)/R(0,T)) is negative. The temperature dependence of MR has a peak-like form with a maximum of the MR amplitude (AMR=max|MR(T)|) in the vicinity of $T_m$ near the point of the maximum of dR/dT. The relative width ($\Delta T/T_m$) of the MR(T) curve is determined at the level 0.5 of AMR, see insert of figure 5b. $\Delta T/T_m$ as function of oxygen index has a minimum for films with optimum oxygen composition, see figure 5b. A variation in both directions of film oxygen content causes an increase of the relative width $\Delta T/T_m$. Such a behavior indicates an increase of structural and spin disorder in the system. The minimum value of AMR in the dependence on $\delta$ corresponds to oxygen excessive composition, and monotonically increases with the increase of the oxygen index. I.e. the MR amplitude increases with a decrease of the concentration of free carriers. We can conclude, that the increase of the MR amplitude reflects the relative increase of the interaction energy with external MF in a comparison with $kT_c$ (k – is Boltzmann constant).

## 3. Discussion

According to the popular crystal-chemistry approach to a description of point defects, the existence of interstitial atoms in crystals with cubic perovskite structure is impossible in equilibrium state [6,11,12]. This is a consequence of the close-packed arrangement of atoms and relates to both cation and anion sublattices. On the other hand, the perovskite structure allows an existence of numerous anion and cation vacancies. When at a non-equilibrium condition excessive oxygen atom is appeared inside crystal, then the scenario of transition to quasi equilibrium state of crystal presupposes the building of addition cell with de-enrichment by atom the neighboring cells. As a result, an excessive oxygen content of the crystal transforms to a cation deficient one.

In our experiments the LSMO films immediately after growth are characterized by an oxygen-deficient composition ($\delta$>0.06). A variation of oxygen content during annealing occurs by means of direct transition of atoms into the neighboring vacancies by the Schottky mechanism. This mechanism provides a variation of the film oxygen content in both directions at relative low temperature. After increase of the oxygen content, the anion flow into the crystal decreases as function of time, as a result of the increase a quantity and length of filled atom chains. This scenario easily explains the $\delta(t)$ saturation at fixed annealing temperature, see figure 1b, as well as an increase of a anion diffusion flow after the stepwise increase of annealing temperature.

For our films, oxygen-excessive compositions were created at annealing temperatures equal to 1170 K in oxygen atmosphere with pressure 1 atm. After annealing, the film structure was cooled down to room temperature relatively slowly (during 1 h), i.e. without quenching the films. Their transport characteristics indicate the existence of oxygen-excessive compositions. The stability of the films state was demonstrated by the following thermal treatment procedures during 4 h at 1170 K and normal pressure of air atmosphere. The transition from an oxygen-excessive composition to oxygen-deficient one occurred after decrease of air pressure down to $10^{-2}$ atm.

According to phase diagrams for powder samples of $La_{0.8}Sr_{0.2}MnO_{3-\delta}$ [6] oxygen excessive compositions can be reached at higher temperatures 1270-1470 K by using relatively small partial oxygen pressures of $10^{-2}$ atm. For our films, the optimum annealing temperature is 1170 K, this condition provides self-organization of a high quality crystal structure [3]. As it was shown, the increase of the annealing temperature up to 1400 K causes a decomposition of the film material in contrast to powder samples.

What is the optimum oxygen content in the film? We have revealed that a maximum temperature $T_m$ exists as function of oxygen index, and have considered this peculiarity as indicator of a minimum amount of defects and distorted lattice cells in the film structure. The variation of the maximum $T_m$ value in different LSMO films depends on the concentration of nonremovable structural defects. One representative of such defects is, first of all, a deviation from stoichiometric composition on Mn.

The presence of defective crystal cells in the film is the reason of a pronounced hysteresis of the R(T) dependence measured in the regime of monotonic variation of temperature. The absence of this hysteresis in films with optimum oxidation confirms our consideration. In principle, our conclusions support the above mentioned approach for a description of point defects formation and adjust the specific phase diagrams [6] for films.

The transport characteristics of LSMO films with optimum oxygen content correspond well with the model of double exchange interaction in the FM state. In the PM region, the temperature dependence of the film conductivity is practically absent. This proves the constant concentration of free charge carriers. A small mobility of charge carriers is the result of both the strong dependence from spin fluctuations scattering and their polaron nature, which increases the electron effective mass up to 5,6 $m_e$ [13].

We have shown, that the statistics of free charge carriers of LSMO film is in good agreement with the equation $z=x-2\delta$ for oxygen excessive compositions. In figure 6 the scheme of film conductivity variation as function of oxygen index is presented for two fixed temperatures 77 and 360 K and for varying temperature $T_m$. The positive values of the oxygen index were determined via the diagrams of [7]. For $\delta<0$, the symmetry of the $\delta(T_m)$ dependence was accepted by us. As it is seen in figure 6, the variation of the film conductivity is relatively small in the region of $\delta<0.046$. This fact reflects a relatively high concentration of free carriers, which is enough for action of double exchange interaction mechanism. In this region the values of σ exceed the threshold value for metal conductance under Mott criterion [4].

In the region $\delta>0.046$ the dependence of σ(δ) is more sharp. The low temperature branch of the σ(δ) dependence reflects a decrease of charge carriers concentration and worsening of the magnetic order with increase of δ. The divergence between the two branches (for 360 K and $T_m$) of this dependence reflects a thermal activation process of free carriers from the localized states which are formed by oxygen vacancies.

An additional parameter for the general estimation of LSMO film quality is the relation between the maximum and minimum values of resistance in the temperature range 77-400 K, see insert of figure 6 [4]. Relatively low values indicate large densities of point defects in the film. In particular, films with excessive compositions of Mn are characterized by values of 3-5 [5].

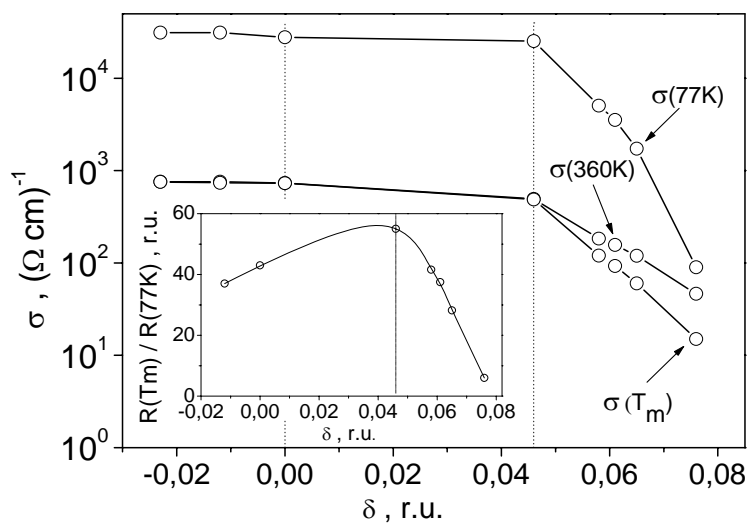

**Figure 6.** The scheme of the film conductivity variation as function of oxygen index at two fixed temperatures (77 and 360K) and at $T=T_m$. Insert: The relation of maximum and minimum values of resistance $R(T_m)/R(77K)$ in the temperature range 77-400 K.

The next peculiarity of the films with an excessive composition of Mn is the temperature dependence of the resistance with an additional low temperature minimum. This peculiarity was explained by a clustered structure in the film and the effect of a Coulomb blockade [5]. Curve 3 of figure 1 demonstrates the same type of R(T) temperature dependence, but the temperature of the resistance minimum in our case is higher, about 150 K. In our case the reason for the R(T) minimum appearance might be caused by a different type of phase separation [14]. An indicator of the appearance of cluster structure in strong defectiveness LSMO films is a decrease of $\sigma(T_m)$ below threshold value $\sigma_{cl} \approx 100$ $(\Omega \cdot cm)^{-1}$, see hierarchical scheme [4].